# Suspended single mode microdisk lasers

*Wanwoo Noh, Matthieu Dupré, Abdoulaye Ndao, Ashok Kodigala, and Boubacar Kanté\**

Department of Electrical and Computer Engineering, University of California San Diego, La Jolla, California, 92093-0407, USA

\* Email: bkante@ucsd.edu

ABSTRACT

Miniature semiconductor lasers have attracted a large amount of interest owing to their potential as highly integrated components in photonic circuits or in sensors. Particularly, microdisk lasers exploiting whispering gallery modes have been regarded as an important candidate because of their relatively small footprint and low threshold. However, it has been challenging for microdisk to operate under single mode operation and to lase in a preselected mode. We report subwavelength microdisk resonators suspended in air with connecting bridges and propose a simple method using the number and symmetry of bridges to enhance or reduce wave confinement in the whispering gallery cavity. Moreover, a suitable choice of bridges increases the quality factor of microdisks compared to microdisks resonator without bridges. Using this method, we demonstrate single mode lasing of preselected modes at telecommunication wavelength.

Over the past few years, small-scale lasers have attracted a widespread attention with potential applications in several areas including health,[1,2] defense,[3] and quality control,[4] to name a few. Recent advances made in photonics, both in understanding physical phenomena[5,6] and in controlling fabrication processes,[7,8] have contributed to improved lasers robustness[9] and size.[10,11]

A semiconductor laser having small modal volume can be designed by implementing various cavities such as Bragg mirror,[12–14] microdisk,[15–21] photonic crystal,[22–25] metallo-dielectric resonator,[26,27] and plasmonic cavity.[28,29] Especially, microdisk lasers using high quality factor whispering gallery mode (WGM) is a promising candidate owing to its low threshold, small modal volume, and simplicity of fabrication. The use of WGMs as lasers is of interest because of their small footprint, their high spectral sensitivity, and their potential for on-chip integration.[30] To exploit the properties of WGMs, a microdisk resonator with an underlying post using undercut etching method has been investigated in many previous studies.[15–21] However, the underlying post used for mechanical support may hamper the intrinsic field profile of WGMs and does not contribute to mode selection.[18]

Another challenge of laser operation is to maintain single mode lasing for specific modes due to the high number of closely spaced WGMs. In addition, any WGMs in the azimuthally symmetric cavity such as cylinder fundamentally has two-fold degeneracy,[30] which brings deleterious effects such as undesirable mode competition and consequent gain saturation, and mode hopping when imperfection appears in practice.[31] To date, single mode lasing has been enabled by breaking degeneracy and suppressing undesirable modes, either by inserting additional features in the resonator such as grating[32] and groove[33] or by breaking parity-time symmetry.[34,35]

Here, we present simple microdisk lasers whose cavity consist of a cylindrical resonator and bridges for self-suspension symmetrically placed at every $2\pi/N$ where N is the number of bridges. The freestanding platform provides higher mode confinement compared to suspended microdisks with a post[15–21] due to the improved index contrast between microdisks and the surrounding medium (air). We also find that bridges not only provide mechanical stability but also maintain or reduce the quality factor of WGMs depending on their configuration. Interestingly, the quality factor of WGMs is enhanced when the number of bridges matches the number of antinodes of WGMs. By optimizing the configuration of bridges, we experimentally demonstrate single-mode lasing at telecommunication wavelength for various WGMs.

RESULTS AND DISCUSSION

Figure 1a illustrates the scanning electron microscope (SEM) image of a fabricated structure with 4 bridges. A cylindrical InGaAsP multiple quantum wells (MQWs) resonator is self-suspended. The suspended resonator is optically pumped (purple beam) and the emission from the device is measured for resonators of different radii and number of bridges. The fabrication process mainly consists of two steps. First, high-resolution pattern definition through e-beam lithography followed by dry etching and second, removal of the bottom sacrificial InP layer by wet etching for suspension.[36] The dimensions of the resonator and bridges are defined in the first step.

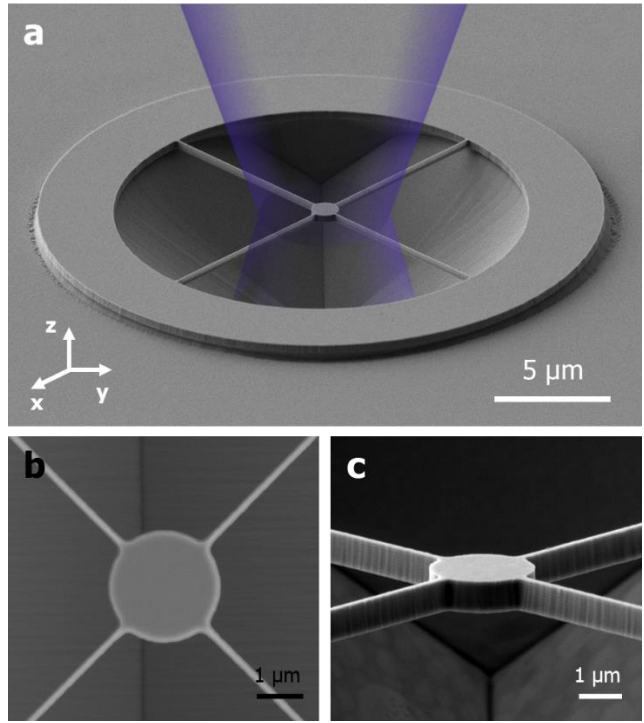

**Figure 1.** Self-suspended microdisk resonator with four bridges. (a) Tilted SEM micrograph of a self-suspended device. The purple beam represents the optical pump beam. (b) Zoom-in top and (c) tilted view of SEM images of the suspended resonator.

Experimental details are presented in METHODS. The zoom-in SEM images, shown in Figure 1b and c, are used to compare the geometrical parameters of the samples to the theoretical specifications.

Modes of the structures are first investigated in the passive structure without bridges using finite element method. Figure 2a illustrates a simple cylinder with a thickness of 300 nm, modeled in the numerical simulation using a refractive index of n=3.45. WGMs are labeled using three indices, the radial index ($l$), the azimuthal index ($m$) and the slab index ($p$). $l$-1, $m$, and $p$-1 represent the number of radial, azimuthal, and z-direction nodes, respectively.[37] TE ($H_z$ dominant) and TM ($E_z$ dominant) WGMs have incomparably higher quality factors than any other Mie mode in the microdisk.[15] Based on our thin disk configuration, the TE modes labeled by $l$, and $p=1$ are predominant in terms of the quality factor.[38] Therefore, we consider only TE WGMs without any radial or slab node ($l$, $p=1$). Figure 2c and d present respectively the resonant wavelength and quality factor for four TE WGMs as a function of their radius. The wavelength range (from 1100 nm to 1700 nm) corresponds to the emission spectrum of InGaAsP MQWs. The normalized electric field and magnetic field distribution of each mode in the middle of the cylinder (z-axis) clearly identifies TE WGMs as shown in Figure 2b. In the gain bandwidth, TE WGMs with $m=3$ to 6 appear and the resonant wavelength of each mode naturally increases as the radius increases (Figure 2c). For a given radius of the cylinder, WGMs with higher $m$ are at a higher frequency. As $m$ increases, the quality factor of WGMs also increases because higher order modes couple less efficiently to the far-field.

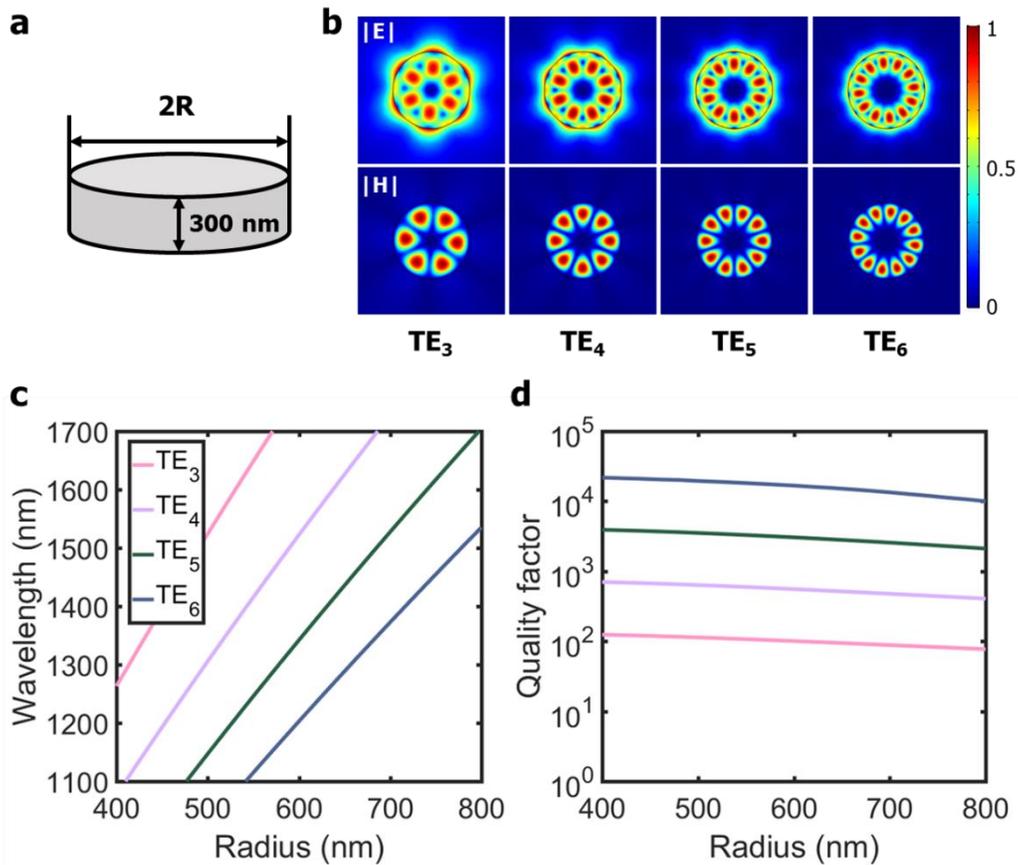

**Figure 2.** Modes of a cylindrical dielectric resonator. (a) Cylinder resonator with characteristic dimensions. (b) Normalized distribution of electric and magnetic field magnitude of WGMs with azimuthal index m=3 to 6. (c) Resonant wavelength and (d) quality factor of four WGMs as a function of the radius of the cylinder.

We now investigate the effect of bridges. They play a role in both the self-suspension of the cavity and the selection of the WGMs. In the simple cylindrical configuration (without bridges), as illustrated in Figure 2b, the WGMs have two notable characteristics in terms of the field distribution. First, the mode is confined around the perimeter of a cylinder. Second, the field pattern has $2m$-fold rotational symmetry. For instance, the TE$_4$ ($l=1$, $m=4$, $p=1$) mode has 8-fold rotational symmetry. Figure 3 illustrates the field distribution of the TE$_4$ mode with a different number of bridges. When the bridges maintain the spatial symmetry of the WGM, the double-degenerate TE$_4$ mode split into two modes due to the spatial perturbation. One of the modes is maintained with the same field profile with a similar or slightly increased quality factor. The second mode does not keep the same field profile and its quality factor is decreased. The lower quality factor modes can be disregarded in considering lasing modes as it will not favorably compete for gain. Cases corresponding to 0, 2, 4, and 8 bridges are presented in Figure 3a. On the other hand, as shown in Figure 3b, if the number of bridges breaks the 8-fold symmetry of TE$_4$ mode pattern, bridges deform the field distribution compared to the simple cylinder case, and this significantly decreases the quality factor from 549 to less than 95. If bridges are placed at every |E| antinodes of TE WGM, the quality factor increases from 549 to 749 as clearly seen in the case of 8 bridges in Figure 3a. Considering the TE$_5$ mode, the quality factor is increased by up to 55% from 2568 (0 bridge) to 3981 (10 bridges).

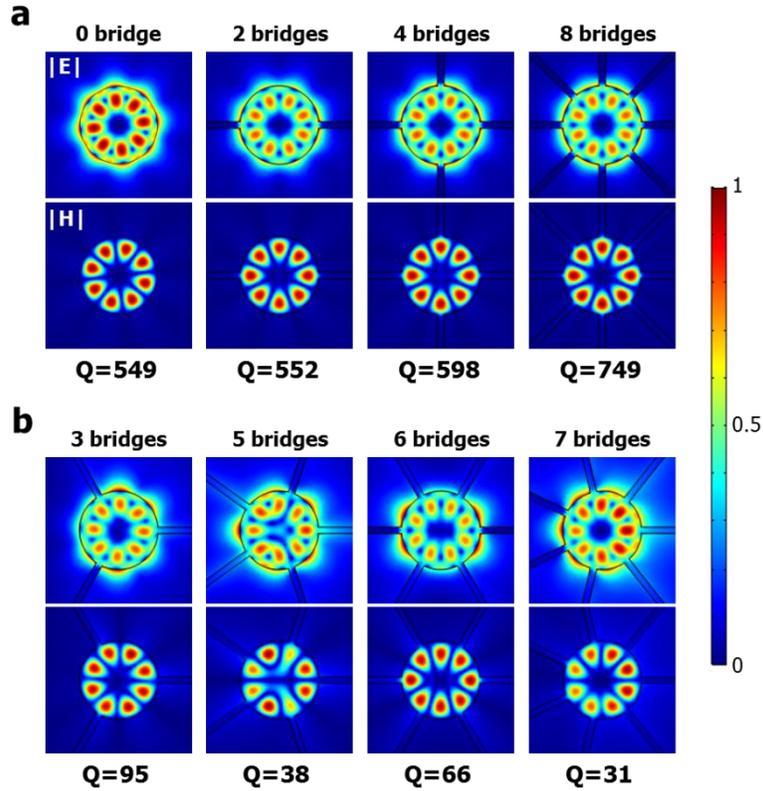

**Figure 3.** Effect of bridges on the field profile of the WGM. Normalized electric and magnetic field distribution of TE$_4$ mode for a cylindrical resonator of radius R=610 nm with the number of bridge matching the azimuthal index of the mode (0, 2, 4, and 8 bridges) (a) mismatching the azimuthal index of the mode (3, 5, 6, and 7 bridges) (b). The quality factor Q of modes is indicated.

To experimentally compare the responses of resonators in different bridge configurations, their normalized photoluminescence as a function of both the radius and the wavelength in the cases of 2 and 4 bridges is measured with the same input power (Figure 4). Each column of Figure 4a and b at given radius is a spectrum measured from a device. The radius of suspended microdisks is varied from 400 nm to 800 nm with a step of 10 nm and total 82 devices are measured for Figure 4a and b. Experimental details are presented in METHODS. Insets represent top view SEM images of one of the measured resonators. It should be noted that 2 bridges placed at 180° from each other do not change the mode pattern of any WGM as they all have at least two nodes. Therefore, any TE WGM with *m* from 3 to 6 clearly appears in Figure 4a. WGMs with larger *m* have higher quality factor and thus are likely to lase if they are in the gain bandwidth of InGaAsP. As a result, with cylinders of larger radius, higher order WGMs first start to lase. In the case of 4 bridges, however, the $TE_3$ and the $TE_5$ mode are significantly suppressed, while $TE_4$ and $TE_6$ mode are protected. As explained, this stems from the fact that the configuration of bridges significantly affects the field profile of WGMs depending on *m*. In the non-lasing regime, not only high-Q TE WGMs are observed in Figure 4, but TM WGMs also appear. However, TM WGMs have lower quality compared to TE WGMs with the same azimuthal index and the difference of quality factor increases for a larger radius of cylinders.

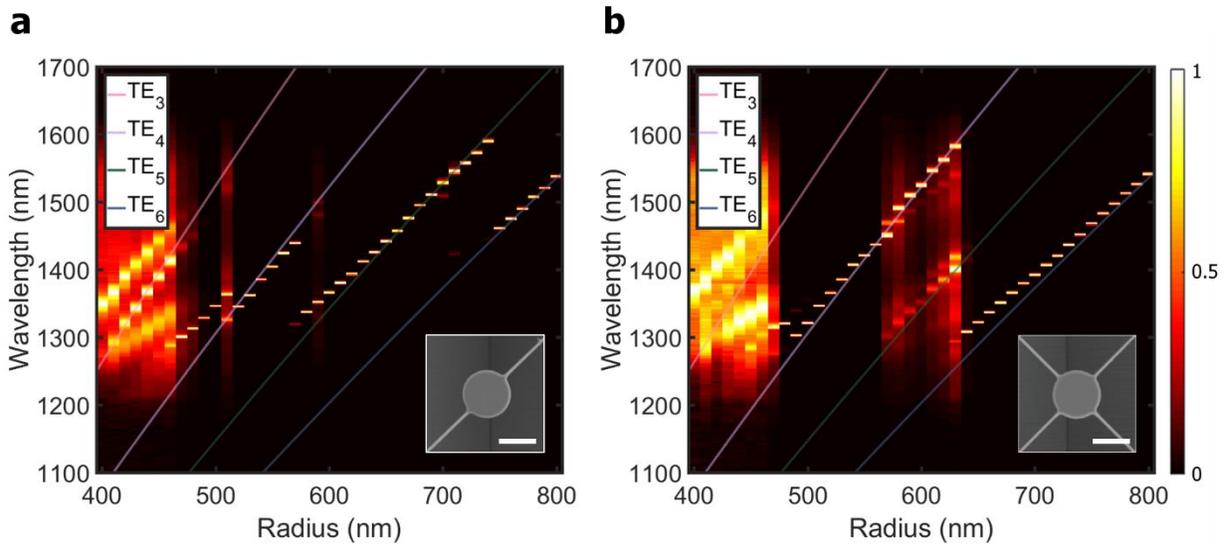

**Figure 4.** Measured and normalized output power from suspended resonators as a function the wavelength and the radius of resonators with 2 bridges (a) and with 4 bridges (b). Resonant wavelength of four TE WGMs (m = 3 to 6) from Figure 2c is also presented to compare experimental results with the numerical simulation. Insets show SEM images of measured resonators with 1 μm scale bars. Lines are eyes guide and clearly show how the measured emission wavelength changes with the radius.

Using the approach above, we demonstrate single-mode microdisk lasers by choosing a suitable number of bridges at the desired wavelength. Figure 5a and b represent normalized emission spectra of devices with radii of 610 and 710 nm, respectively, designed to have $TE_4$ and $TE_5$ mode with the resonant wavelength at 1.55 μm (Figure 2c). As discussed, a WGM has the highest quality factor when the number of bridges is the same as the number of antinodes or 2*m*. These resonators are designed to have 8 and 10 bridges to have an optimal quality factor. For comparison, the 2-bridge case is selected as a reference to illustrate the effect of bridges on the lasing mode. Red curves in Figure 5a and c represent the optimized bridge configurations, while blue curves correspond to the 2-bridge cases. The position of bridges is uniformly distributed but avoid the crystal plane of InP because of possible difficulties in removing the sacrificial InP layer in wet-etching. Insets show top-view SEM images of the measured device with a different number of bridges. In figure 5a, the 2-bridge case shows lasing action at the undesired wavelength (~1.4 μm) due to the appearance of a higher order WGM ($TE_5$). However, with 8 bridges, single-mode lasing is observed at telecommunication wavelength due to of a 710 nm radius (Figure 5b), the device shows single mode lasing at 1.55 μm as well, whereas suppression of other existing WGMs with different azimuthal orders m. In the case of the resonator microdisks with 2 bridges show lasing actions at different and several wavelengths. Since the cylindrical resonator with 2 bridges has the same modal response as a pure WGMs, the 2-bridge resonator exhibits other high-Q WGMs which parasitically contribute to multimode lasing. We also measured light-light characteristics and threshold powers of the two laser devices operating at 1.55 μm. Figure 5b and d correspond to 610 and 710 nm cylindrical resonators with optimized bridge configurations. A clear transition from spontaneous emission to stimulated emission is observed in both cases, confirming that reported devices are indeed lasers.

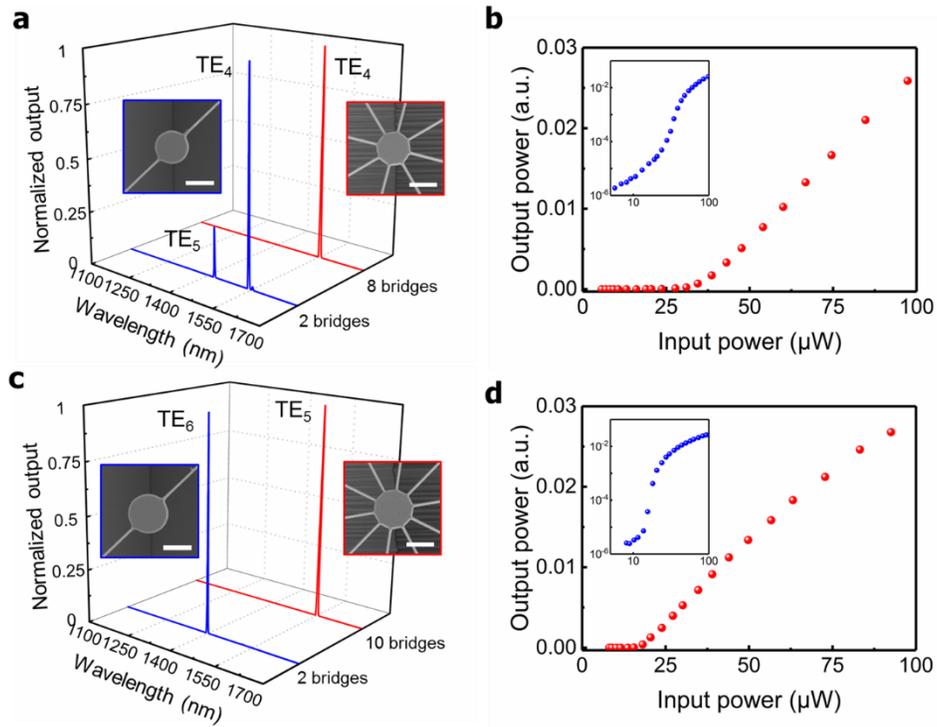

**Figure 5.** Normalized emission spectra measured from self-suspended microdisks having radius of (a) 610 nm and (c) 710 nm designed to support the $TE_4$ and $TE_5$ mode, respectively, at 1.55 μm wavelength. The blue line corresponds to the 2-bridge case used as a reference, while red line corresponds to the optimized bridge configurations with 8 and 10 bridges. Insets show SEM images of measured devices and the scale bars represent 1 μm. Light-light curves clearly show a threshold behavior of the emitted power as a function of pump power, i.e., lasing from optimized devices exploiting the $TE_4$ (c) and the $TE_5$ mode (d). Insets represent log-scale light-light plot.

CONCLUSION

We proposed and demonstrated self-suspended microdisk lasers via bridges. The bridges play two important roles. They facilitate the mechanical stability of the devices and also provide higher wave confinement in the cavity due to the large refractive index contrast with the environment. More importantly, the bridges can be used to spatially control the field profile of WGMs and therefore to control their quality factor by protecting or breaking their fundamental $2m$-fold rotational symmetry. By controlling the quality factor of WGMs from the configuration of the optimal bridges, we have demonstrated single-mode lasing devices that exploit the $TE_4$ and the $TE_5$ WGMs at telecommunication wavelength. Our approach can be used to engineer WGMs of arbitrary order and is expected to serve as a useful scheme toward single mode lasing in various platforms.

METHODS

Fabrication of suspended resonator. We follow typical nanofabrication steps to fabricate a single resonator suspended in air. A 300nm-thick InGaAsP MQWs are epitaxially grown on InP wafer. 10 nm wells and 20 nm barriers are alternating 9 times and the composition ratio between III and IV group are designed to have most gain around near-IR (~1.5 μm). The fabrication steps mainly consist of two steps, which are patterning and suspension. Fabrication begins with cleaning sample through sonication in order of acetone, then isopropyl alcohol and finally distilled water for 10 min each. The negative tone hydrogen silsesquioxane (HSQ) e-beam resist is spin-coated with a 200 nm thickness. The high-resolution e-beam lithography (Vistec EBPG5200 writer) defines a highly fine pattern of a cylindrical resonator and its supporting bridges. After developing, we dry etch the InGaAsP and InP layers to pattern the resonator and the bridges. We use reactive ion etching with plasma of $H_2$, $CH_4$, and Ar for 550 s to etch about 500 nm. Etch depth should be more than 300 nm, which is the thickness of InGaAsP for further suspending process. After O$_2$ plasma cleaning for 5 min to remove the residuals and removing e-beam resist by buffered oxide etch (BOE) process, a 1.5 μm-thick NR9-1500PY negative tone photoresist is spin-coated, followed by light exposure for 20 s using a mask aligner (Karl Suss MA-6). This process creates an opening to protect the InP layer from unnecessary removal. We suspend the resonator by selectively wet-etching the sacrificial layer with an HCl solution, which does not react with InP but with InGaAsP. After putting the sample in PG remover at 80 °C overnight to remove photoresist, our device is finalized.

Optical characterization. Emission spectra are measured using a micro-photoluminescence setup. To optically pump suspended resonator, 1064 nm pulsed laser with 12 nm pulse width and 215 kHz repetition rate is used at room temperature. Pumping beam is focused onto the sample after passing through the microscope objective with the numerical aperture of 0.4. Emitted light from the sample propagates through the double 4-f lens system and finally focused either on InGaAs detector or on near-infrared camera by using a flip mirror. The grating-based monochromator enables to measure photoluminescence by the detector at a discrete wavelength with a resolution of 0.33 nm.